\documentclass[apps,pra,twocolumn,armsfonts,amssymb,armsmath,showpacs,floatfix,nofootinbib,groupedaddress,superscriptaddress,citesort]{revtex4}
\usepackage{mathrsfs}
\usepackage{amsfonts}
\usepackage{amstext}
\usepackage{amsmath}
\usepackage{amssymb}
\usepackage{float}
\usepackage[usenames]{xcolor}
\usepackage[dvips]{graphicx}
\usepackage[colorlinks, citecolor=blue]{hyperref}

\def\ra{\rangle}
\def\la{\langle}

\def\no{\nonumber}
\def\bea{\begin{eqnarray}}
\def\eea{\end{eqnarray}}
\def\be{\begin{equation}}
\def\ee{\end{equation}}
\def\c{\cos r}
\def\s{\sin r}
\def\cc{\cos^2 r}
\def\ss{\sin^2 r}

\begin{document}
\title{Entropic uncertainty relations under the relativistic motion}

\author{Jun Feng}
\email{tsunfeng@iphy.ac.cn}
\affiliation{Beijing National Laboratory for Condensed Matter Physics,
Institute of Physics, Chinese Academy of Sciences, Beijing 100190, P. R. China}
\author{Yao-Zhong Zhang}
\affiliation{School of Mathematics and Physics, The University of Queensland, Brisbane, Qld 4072, Australia}
\author{Mark D. Gould}
\affiliation{School of Mathematics and Physics, The University of Queensland, Brisbane, Qld 4072, Australia}
\author{Heng Fan}
\email{hfan@iphy.ac.cn}
\affiliation{Beijing National Laboratory for Condensed Matter Physics,
Institute of Physics, Chinese Academy of Sciences, Beijing 100190, P. R. China}

\begin{abstract}

The uncertainty principle bounds our ability to simultaneously predict two incompatible observables of a quantum particle. Assisted by a quantum memory to store the particle, this uncertainty could be reduced and quantified by a new Entropic Uncertainty Relation (EUR). In this Letter, we explore how the relativistic motion of the system would affect the EUR in two sample scenarios. First, we show that the Unruh effect of an accelerating particle would surely increase the uncertainty if the system and particle entangled initially. On the other hand, the entanglement could be generated from nonuniform motion once the Unruh decoherence is prevented by utilizing the cavity. We show that, in a uncertainty game between an inertial cavity and a nonuniformly accelerated one, the uncertainty evolves periodically with respect to the duration of acceleration segment. Therefore, with properly chosen cavity parameters, the uncertainty bound could be protected. Implications of our results for gravitation are also discussed.\\

\end{abstract}
\pacs{03.65.Ta, 03.65.Yz, 04.62.+v}
\maketitle

\section{Introduction}
\label{1}

The distinguishability of quantum theory from classical theory is formulated by Heisenberg's uncertainty relation \cite{H1}, which bounds the uncertainties about the measurement outcomes of two incompatible observables of a quantum particle. Nowadays, this principle has been recast, in an information-theoretical framework, with the uncertainty quantified by entropic measures \cite{H2,H3}. In particular, it states \cite{H4,H5,H6} that for any observables $Q$ and $R$
\be
H(Q)+H(R)\geqslant\log_2\frac{1}{c}
\label{eur1}
\ee
where $H=-\sum_xp(x)\log_2p(x)$ denotes the Shannon entropy for the probability distribution of the measurement outcomes, $c=\mbox{max}_{i,j}|\la a_i| b_j\ra|^2$ represents the overlap between observables $Q$ and $R$ with $|a_i\ra$ and $|b_j\ra$ the corresponding eigenvectors, respectively. Since $c$ does not depend on specific states to be measured, the right-hand side (RHS) of inequality (\ref{eur1}) provides a fixed lower bound and more general framework of quantifying uncertainty than the standard deviations. 

However, using previously determined quantum information about the measured system, the above uncertainty  bound could be violated. This dramatically gives a stronger Entropic Uncertainty Relation (EUR) which has been proved recently \cite{EUR1,EUR2}, followed by several experimentally confirmation \cite{EUR3,EUR4}. The new relation can be illustrated by the uncertainty game between two players Alice and Bob, where Bob prepares a particle in a quantum state of his choosing and sends it to Alice, who then carries out one of the two measurements and announces her choice to Bob. If the particle of interest ($A$) initially entangles with another particle, acting as a quantum memory ($B$) carried by the observer, the new EUR is expressed as
\be
S(Q|B)+S(R|B)\geqslant\log_2\frac{1}{c}+S(A|B)
\label{eur2}
\ee
where $S(A|B)=S(\rho_{AB})-S(\rho_{B})$ is the conditional von Neumann entropy. After the quantum system $A$ is measured by $X$ with $X\in(Q,R)$, the post measurement state is $\rho_{XB}=\sum_x(\Pi_x\otimes\mathbf{1})\rho_{AB}(\Pi_x\otimes\mathbf{1})$, where $\Pi_x=|\psi_x\ra\la\psi_x|$ and $\{|\psi_x\ra\}$ are the eigenstates of the observable $X$. In the extreme case, when $A$ and $B$ are maximally entangled, it is able to predict the outcomes precisely. On the other hand, if $A$ and $B$ are not entangled, the bound in (\ref{eur1}) is recovered. The generalization of the quantum-memory-assisted EUR (\ref{eur2}) to R\'enyi entropy has also been given \cite{EUR5,EUR6}. Other studies from various views can be found in \cite{EUR7,EUR8,EUR9}.

In a realistic regime, quantum systems inevitably suffer a decoherence or dissipation resulting from the interaction between the systems and the environment. Therefore, it is important to investigate the influence of environment decoherence on quantum-memory-assisted EUR, which establish nontrivial dependence on noisy channel \cite{APP1}.

In this Letter, we explore the EUR (\ref{eur2}) under the decoherence rooting in the relativistic motion of quantum systems. Besides the measurement the observer performed, we find the uncertainty bound in (\ref{eur2}) \emph{does} depend on the motion state of the system. In the first scenario, we consider the bipartite system in which Bob is under an uniformly acceleration relative to his inertial partner Alice. Since they differ in their description of a given quantum state due to the celebrated Unruh effect, the concept of entropy should be observer-dependent in this situation \cite{ADD1}, consequently,  implying a nontrivial relativistic modification to the conditional entropy in EUR (\ref{eur2}).  In our uncertainty game, Bob send a qubit $A$, constructed from the free fermionic mode, to Alice. We show that the quantum information stored initially in $B$ would be degraded, therefore resulting an inevitably increase of the uncertainty on the outcome of measurements carried by Alice. In the limit of infinite acceleration, the uncertainty reach a finite maximal value. This phenomenon is essentially different from that reported in \cite{APP1} where entanglement transfers between the quantum system and its environment. In our scenario, all field modes beyond Bob's acceleration horizon have to be abandoned, therefore the quantum information is completely damaged \cite{RQI2,RQI3}. 

By utilizing the localized modes in cavity, the Unruh decoherence could be prevented in either Minkowskian or curved spacetime \cite{CAVITY1,CAVITY2}. Nevertheless, the entanglement between the cavity modes could still be generated by the nonuniform motion of the single cavity \cite{BOX1,BOX2,BOX3}. In our second scenario, we perform the uncertainty game between Alice and Bob, with the particle (qubit) $A$ and quantum memory $B$, each restricted in a rigid cavity, maximally entangled initially. Once Bob's cavity moves along a trajectory consisted of inertial and then uniformly accelerated segments, we show that the uncertainty would evolve periodically with respect to the duration of acceleration segment. With properly chosen cavity parameters, the uncertainty bound could be protected. We show explicitly how the uncertainty bound of (\ref{eur2}) depends on the dynamics of the cavity. Moreover, our analysis admits a low-acceleration approximation, which means the results in the Letter are comparable with possible real experiments.

\section{Quantum-memory-assisted EUR in noninertial system}
\label{2}

We first recall some backgrounds on Unruh effect from the quantum information view \cite{RQI1}. In Minkowski space, the field modes in the view of an observer with the acceleration $a$ should be described by Rindler coordinates, dividing the whole spacetime into left and right Rindler wedges by acceleration horizon. Since the field modes restricted in different wedges cannot be causally connected, the information loss for the accelerated observer results a thermal bath. In particular, considering fermionic field with few degrees for simplicity, the vacuum state $|0\ra=\bigotimes_{k}|0_k\ra$ and its excitation $|1\ra=\bigotimes_{k}|1_k\ra$  in Minkowskian coordinates can be expressed as 
\bea
|0_k\ra&=&\c |0_k\ra_{\mbox{\tiny I}}|0_k\ra_{\mbox{\tiny II}}+\s |1_k\ra_{\mbox{\tiny I}}|1_k\ra_{\mbox{\tiny II}}\no\\
|1_k\ra&=& |1_k\ra_{\mbox{\tiny I}}|0_k\ra_{\mbox{\tiny II}}
\label{bog1}
\eea
in terms of modes in different Rindler wedges I and II, where $\tan r=e^{-\pi\omega/a}$. It should be noted that \cite{SMA1}, the so-called Single-Mode Approximation (SMA) has be imposed, which however is not correct for general states. Indeed, the Bogoliubov transformation (\ref{bog1}) corresponds a particular choice of so-called Unruh mode, which are symmetric over both left and right Rindler wedges. Nevertheless, this approximation could work when only the main contributing mode of a well-chosen wave packet is considered \cite{SMA2,SMA3}.

As mentioned before, in the uncertainty game, Bob sends Alice a qubit $A$, initially entangled with another his quantum memory $B$. After Alice measuring either $Q$ or $R$ and broadcasting her measurement choice, Bob needs to minimize his uncertainty about Alice’s measurement outcome.

To investigate the influence of Unruh effect on this game, we consider the bipartite system of Alice and Bob, initially being static and sharing a Bell-diagonal state
\be
\rho_{AB}=\frac{1}{4}(\mathbf{1}^A\otimes\mathbf{1}^B+\sum_{i=1}^{3}c_i\sigma_i^A\otimes\sigma_i^B)
\label{bell}
\ee
where $\sigma_i$ are Pauli matrix and the coefficients $0\leqslant|c_i|\leqslant1$. In particular, this state reduces to maximally entangled states (Bell-basis) if $|c_1|=|c_2|=|c_3|=1$. Following \cite{RQI3}, we assume all states of Alice are constructed by the field mode $s$ which only her detector can detect, while Bob's states can be truncated from $k$ field modes to which his detector is sensitive. With this convention in mind, the superscripts $A$ and $B$ would have double meaning as particular observer and the field modes he/she sensitive. After their coincidence, we assume one observer move with an uniform acceleration while another one maintain static. Since the accelerating observer cannot access the modes beyond his/her horizon, the lost information reduces the entanglement between $A$ and $B$, therefore change the uncertainty in (\ref{eur2}). In following, we discuss the scheme with Bob (or Alice) being under an acceleration, respectively.

In our uncertainty game, after sending Alice the qubit $A$ maximally entangled with quantum memory $B$, Bob begins to move with an acceleration $a$. The qubit $B$ in (\ref{bell}) should be transformed according to (\ref{bog1}) and traced over all degrees in region II, in particular, we have 
\bea
\rho_{AI}&=&\frac{1}{4}\bigg(\mathbf{1}^A\otimes\big[\cc|0\ra_I\la0|+(\ss+1)|1\ra_I\la1|\big]^B\no\\
&+&\sum_{k=1}^2c_k\c\sigma_k^A\otimes\sigma_{I,k}^B+c_3\cc\sigma_3^A\otimes\sigma_{I,3}^B\bigg)
\label{bacc}
\eea
where $\sigma_{I,1}^B=|0\ra_I\la1|+|1\ra_I\la0|$, $\sigma_{I,2}^B=-i|0\ra_I\la1|+i|1\ra_I\la0|$ and $\sigma_{I,3}^B=|0\ra_I\la0|-|1\ra_I\la1|$ are Pauli matrix restricted in region I.

The qubit $A$ is measured by one of the Pauli operators $\sigma_i$. Since Alice is free from Unruh effect, the measurement outcomes should be independent on the motion state of quantum memory. However,  the conditional von Neumann entropy in (\ref{eur2}) would be changed, since (\ref{bacc}) gives new post measurement states $\sum_x(\Pi_x\otimes\mathbf{1})\rho_{AI}(\Pi_x\otimes\mathbf{1})$, restricted in region I. Employing the eigenstates of Pauli operators $\sigma_i$, we have the post measurement states
\bea
\rho_{\sigma_iI}&=&\frac{1}{4}\bigg(\mathbf{1}^A\otimes\big[\cc|0\ra_I\la0|+(\ss+1)|1\ra_I\la1|\big]^B\no\\
&+&c_i\c\sigma_i^A\otimes\sigma_{I,i}^B\bigg)\qquad\quad(i=1,2)\label{bpm1}\\
\rho_{\sigma_3I}&=&\frac{1}{4}\bigg(\mathbf{1}^A\otimes\big[\cc|0\ra_I\la0|+(\ss+1)|1\ra_I\la1|\big]^B\no\\
&+&c_3\cc\sigma_3^A\otimes\sigma_{I,3}^B\bigg)\label{bpm3}
\label{bacc-iI}
\eea
whose eigenvalues can be easily obtained and give the von Neumann entropy
\bea
S(\rho_{\sigma_iI})&=&H_{\mbox{\tiny bin}}(\frac{4-\sqrt2\lambda_i}{8})+1\qquad\quad(i=1,2)\label{bent1}\\
S(\rho_{\sigma_3I})&=&H'_{\mbox{\tiny bin}}(\lambda_3^+)+H'_{\mbox{\tiny bin}}(\lambda_3^-)\label{bent3}
\eea 
where
\bea
\lambda_i&=&\sqrt{3+4c_i^2+4(c_i^2-1)\cos(2r)+\cos(4r)}\no\\
&&(i=1,2)\no\\
\lambda_3^\pm&=&\frac{(1\pm c_3)\cc}{4}\no
\eea
Here we denote $H_{\mbox{\tiny bin}}(p)=-p\log_2p-(1-p)\log_2(1-p)$ as the binary entropy, and similarly $H'_{\mbox{\tiny bin}}(p)=-p\log_2p-(\frac{1}{2}-p)\log_2(\frac{1}{2}-p)$.

\begin{figure}[hbtp]
\includegraphics[width=.43\textwidth]{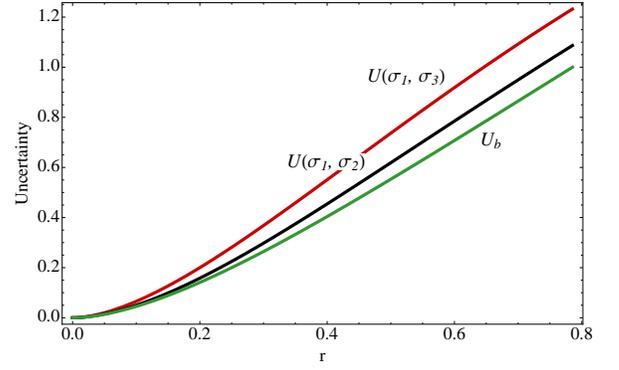}
\caption{The uncertainty of the measurement on observables $\sigma_1$ and $\sigma_3$ (or  $\sigma_2$) depends on the acceleration of Bob. The maximally entangled initial state with $c_1=-c_2=c_3=1$ has been chosen, and $r=\frac{\pi}{4}$ corresponds the limit of infinite acceleration.}
\label{EUR-Bacc}
\end{figure}

Since the accelerated quantum memory is $\rho_I=\mbox{Tr}_{\mbox{\tiny A}}\rho_{AI}$, the associated entropy is also binary $S(\rho_I)=H_{\mbox{\tiny bin}}(\frac{\cc}{2})$. For a particular measurement of $\sigma_1$ and $\sigma_3$ by Alice, we can give the left-hand side (LHS) of (\ref{eur2}), denoted as uncertainty $U$ 
\bea
U(\sigma_1,\sigma_3)&=&H_{\mbox{\tiny bin}}(\frac{4-\sqrt2\lambda_1}{8})-2H_{\mbox{\tiny bin}}(\frac{\cc}{2})\no\\
&+&H'_{\mbox{\tiny bin}}(\lambda_3^+)+H'_{\mbox{\tiny bin}}(\lambda_3^-)+1
\label{lhs1}
\eea
depicted in Fig. \ref{EUR-Bacc} (similar result for the measurement on $\sigma_1$ and $\sigma_2$ is also given). It is clear that the uncertainty grows with respect to larger $r$, representing the acceleration of Bob. This result should not surprise us since the Unruh effect reduces the entanglement between $A$ and $B$, therefore lowers our ability to preciously predict the outcomes of Alice's measurement.

We now investigate the RHS of (\ref{eur2}). Once the measurement choice has been determined, the complementarity $c$ of the observables $\sigma_j$ and $\sigma_k$ is always $1/2$. The conditional entropy now is $S(A|I)=S(\rho_{AI})-S(\rho_I)$. The eigenvalues of (\ref{bacc}) are
\bea
\eta_{1,\pm}&=&\frac{1}{16}[4(1+c_3\cc)\pm\sqrt2\bar{\lambda}_-]\no\\
\eta_{2,\pm}&=&\frac{1}{16}[4(1-c_3\cc)\pm\sqrt2\bar{\lambda}_+]\no
\eea
where $\bar{\lambda}_\pm=\sqrt{3+8(c_1\pm c_2)^2\cc-4\cos(2r)+\cos(4r)}$. Denoting the RHS of (\ref{eur2}) as $U_b$, we have
\be
U_b=-\sum_{i=1,2;
j=\pm}\eta_{i,j}\log_2\eta_{i,j}-H_{\mbox{\tiny bin}}(\frac{\cc}{2})+1
\label{rhs1}
\ee
depending on the acceleration of Bob through the parameter $r=\arctan(e^{\pi\omega/a})$.

As depicted in Fig. \ref{EUR-Bacc} for Bell state with $c_1=-c_2=c_3=1$, in the limit of vanish acceleration, the $A$ and $B$ remain be maximally entangled, $U=U_b=0$, satisfying the EUR (\ref{eur2}). However, as the acceleration of Bob grows, the uncertainty bound $U_b$ is lifted and $U>U_b$. In the infinite acceleration limit ($r\rightarrow\frac{\pi}{4}$), the uncertainty reaches a finite maxima. This means that the state preparation and measurement choice (SPMC) condition for inertial system, $c_i=-c_jc_k\  (i\neq j\neq k)$ \cite{APP1}, will no longer be satisfactory but should be replaced by the involved acceleration-dependent condition $U\equiv U_b$ from (\ref{lhs1}) and (\ref{rhs1}). 

Finally, we briefly discuss an alternative game, in which Alice accelerates after sharing (\ref{bell}) with inertial Bob. Since Alice would suffer the Unruh effect, the local measurements on $X$ should be made in region I, which means a restricted projector $\Pi_I=|\psi_x\ra_I\la\psi_x|$ and $\{|\psi_x\ra\}_I$ are the eigenstates of the observable $X$ in region I. The symmetric character of $A$ and $B$ in Bell-diagonal state guarantees the same post measurement states as (\ref{bpm1})-(\ref{bpm3}) with exchange the superscripts. the new bipartite state is same as (\ref{bacc}) but with exchanged superscripts. Nevertheless, unlike in before scheme, the outcomes of Alice's measurement now have been changed by the Unruh effect. After straightforward calculation, we can show there is no quantitative difference from before scheme with accelerating Bob. Therefore, all results given before are still correct.

\section{Uncertainty game with nonuniform-moving cavity}
\label{3}

We now discuss an alternative scenario in which both Alice and Bob are localized in rigid cavity. While the rigid boundaries of cavity protect the inside observer from the Unruh effect, the relativistic motion of the cavity would still affect the entanglement between the free field modes inside \cite{BOX1,BOX2,BOX3}, therefore leading a motion-dependent uncertainty bound.

For simplicity, it is enough to introduce a model in $(1+1)$-dimensional Minkowski spacetime, where the massless Dirac equation $i\gamma^\mu\partial_\mu\psi=0$ admits left/right Weyl spinor solution with opposite helicity. The cavity with length $L=x_2-x_1$ imposes the Dirichlet conditions on the eigenfunctions $\psi_n(t,x)$ of the Hamiltonian. Once the cavity accelerates, it is convenient to use the Rindler coordinates $(\eta,\chi)$, defined in the wedge $x>|t|$ by $t=\chi\sinh\eta$ and $x=\chi\cosh\eta$, where $0<\chi<\infty$ and $-\infty<\eta<+\infty$. The new orthonormal eigenfunctions $\hat{\psi}_n(\eta,\chi)$ can be derived by solving the massless Dirac equation in Rindler coordinates.

\begin{figure}[hbtp]
\includegraphics[width=.4\textwidth]{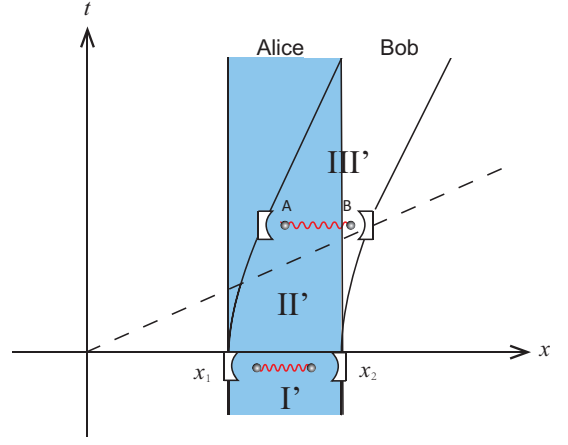}
\caption{The bipartite system of particle A to be measured and the quantum memory B in a cavity. The cavity trajectory of nonuniform motion is initially in the inertial segment I', after the acceleration segment II', returning to the inertial segment III'. In a uncertainty game with Alice and Bob restricted in distinct cavities, the uncertainty bound can be protected. }
\label{space2}
\end{figure}

Consider a cavity trajectory of nonuniform motion shown in Fig. \ref{space2}.  The cavity begins to accelerate at $t=0$, following the Killing vector $\partial_\eta$. The acceleration ends at Rindler time $\eta=\eta_1$, and the duration of the acceleration in proper time measured at the center of the cavity is $\tau_1=\frac{1}{2}(x_1+x_2)\eta_1$. The three segments of the trajectory are referred as I', II' and III'.

In Fock space, the Dirac field can be expanded in quantized eigenfunctions as $\psi=\sum_{n\geqslant0}a_n\psi_n+\sum_{n\leqslant0}b^\dag_n\psi_n$ in segment I'. Similar expansions could be made with $\hat{\psi}_n$ in segment II' and $\tilde{\psi}_n$ in segment III'. The nonvanishing anticommutators in segment I' are $\{a_m,a^\dag_n\}=\{b_m,b_n^\dag\}=\delta_{mn}$, defining the vacuum $a_n|0\ra=b_n|0\ra=0$, similar definitions for other two regions. Any two field modes in distinct regions can be related by Bogoliubov transformations like
\bea
\hat{\psi}_m=\sum_{n}A_{mn}\psi_n&,&\psi_n=\sum_mA_{mn}^*\hat{\psi}_m\no\\
\tilde{\psi}_m=\sum_{n}\mathbb{A}_{mn}\psi_n&,&\psi_n=\sum_m\mathbb{A}_{mn}^*\tilde{\psi}_m
\eea
In the limit of small cavity acceleration, which gives the comparable results for real experiment, these coefficients can be calculated perturbatively. More specific, by introducing the dimensionless parameter $h=2L/(x_1+x_2)$, which is the product of the cavity's length and the acceleration at the center of the cavity, the coefficients can be expanded in a Maclaurin series to $h^2$ order, $A=A^{(0)}+A^{(1)}+A^{(2)}+\mathcal{O}(h^3)$, and similarly $\mathbb{A}=\mathbb{A}^{(0)}+\mathbb{A}^{(1)}+\mathbb{A}^{(2)}+\mathcal{O}(h^3)$. The explicit form of coefficients can be found in \cite{BOX1}.

The Fock vacua $|0\ra$ in segment I' and $|\tilde{0}\ra$ in segment III' can be related by a squeezed operators
$|0\ra=Ne^W|\tilde{0}\ra$, where $N$ is normalization constant and $W=\sum_{p\geqslant0,q\leqslant0}V_{pq}\tilde{a}_p^\dag\tilde{b}_q^\dag$. Denoting the one-particle states in segment III' by $|\tilde{1}_k\ra^+=\tilde{a}_k^\dag|\tilde{0}\ra$ for $k\geqslant0$ and $|\tilde{1}_k\ra^-=\tilde{b}_k^\dag|\tilde{0}\ra$ for $k\leqslant0$, one has \cite{BOX1}
\bea
|0\ra&=&\bigg(1-\frac{1}{2}\sum_{p,q}|V_{pq}|^2\bigg)|\tilde{0}\ra+\sum_{p,q}V_{pq}|\tilde{1}_p\ra^+|\tilde{1}_q\ra^-\no\\
&-&\frac{1}{2}\sum_{p,q,i,j}V_{pq}V_{ij}(1-\delta_{pi})(1-\delta_{qj})\no\\
&&\times|\tilde{1}_p\ra^+|\tilde{1}_i\ra^+|\tilde{1}_q\ra^-|\tilde{1}_j\ra^-+\mathcal{O}(h^3)
\label{vac}
\eea
The one-particle states in segment I' can be derived from $|1_k\ra^+=a_k^\dag|0\ra$.

We now turn to the uncertainty game between Alice and Bob who is protected by the rigid cavity from the possible Unruh effect. Initially, both Alice and Bob in a inertial in-region I' and share the Bell-diagonal state (\ref{bell}), but with the single-particle states $|1_k\ra^+=a_k^\dag|0\ra$. For instance, when $c_1=-c_2=c_3=1$, they share the maximally entangled state
\be
\frac{1}{\sqrt2}(|0_{\hat{k}}\ra^A|0_k\ra^B+|1_{\hat{k}}\ra^{+A}|1_k\ra^{+B}\ra)
\ee
The superscripts $A$ and $B$ refer the particle to be measured and quantum memory in different cavities, and $\hat{k}$ and $k$ are distinct modes sensitive for Alice and Bob respectively.

After their coincidence, Bob's cavity begins accelerating in segment II' and then becomes inertial again in out-region III', while Alice always maintains her inertial motion, see Fig. \ref{space2}. Similar as before,  in the out-region, all Bob's states have been transformed according to (\ref{vac}). Since the only reference mode of Bob is mode $k$, all other modes of Bob should be traced over, therefore inducing the information loss due to the acceleration.

We have the new bipartite state
\be
\rho_{A\tilde{B}}=\frac{1}{4}\bigg(\mathbf{1}^A\otimes\tilde{\mathbf{1}}^B+\sum_{i=1}^3c_i\sigma_i^A\otimes\tilde{\sigma}_i^B\bigg)
\label{cavity1}
\ee
where 
\bea
\tilde{\mathbf{1}}^B&=&(1+F_-)|\tilde{0}_k\ra\la \tilde{0}_k|+(1-F_-)|\tilde{1}_k\ra^{++}\la \tilde{1}_k|\no\\
\tilde{\sigma}_1^B&=&(G_k+\mathbb{A}^{(2)}_{kk})|\tilde{0}_k\ra^+\la \tilde{1}_k|+(G_k+\mathbb{A}^{(2)}_{kk})^*|\tilde{1}_k\ra^+\la \tilde{0}_k|\no\\
\tilde{\sigma}_2^B&=&-i(G_k+\mathbb{A}^{(2)}_{kk})|\tilde{0}_k\ra^+\la \tilde{1}_k|+i(G_k+\mathbb{A}^{(2)}_{kk})^*|\tilde{1}_k\ra^+\la \tilde{0}_k|\no\\
\tilde{\sigma}_3^B&=&(1-F_+)(|\tilde{0}_k\ra\la \tilde{0}_k|-|\tilde{1}_k\ra^{++}\la \tilde{1}_k|)
\label{coef}
\eea
where the phase factor $G_k=\exp[i(k+s)\pi\eta_1/\ln(x_2/x_1)]$ with $s\in[0,1)$ characterizing the self-adjoint extension of the Hamiltonian \cite{PHASE}. All other coefficients are given explicitly in \ref{5}.

Since Alice moves always inertially, the post measurement states  $\sum_x(\Pi_x\otimes\mathbf{1})\rho_{A\tilde{B}}(\Pi_x\otimes\mathbf{1})$ would have similar form as in (\ref{bpm1}) - (\ref{bpm3})
\be
\rho_{\sigma_i\tilde{B}}=\frac{1}{4}\bigg(\mathbf{1}^A\otimes\tilde{\mathbf{1}}^B+c_i\sigma_i^A\otimes\tilde{\sigma}_i^B\bigg)
\ee
admitting the form of X-type matrix, whose eigenvalues can be solved explicitly. Here we list the corresponding von Neumann entropy
\bea
S(\rho_{\sigma_i\tilde{B}})&=&H_{\mbox{\tiny{bin}}}(\frac{1-\tilde{\lambda}_i}{2})+1 \qquad\qquad(i=1,2)\no\\
S(\rho_{\sigma_3\tilde{B}})&=&H'_{\mbox{\tiny{bin}}}(\tilde{\lambda}_3^+)+H'_{\mbox{\tiny{bin}}}(\tilde{\lambda}_3^-)\no
\eea
where
\bea
\tilde{\lambda}_i&=& \sqrt{F_-^2+c_i^2|G_k+\mathbb{A}^{(2)}_{kk}|^2})\qquad\quad(i=1,2)\no\\
\tilde{\lambda}_3^\pm&=&\frac{1-c_3(1-F_+)\pm F_-}{4}\no
\eea
To calculate the conditional von Neumann entropy associated with the chosen measurement, we give the entropy of the reduced density matrix $\rho_{\tilde{B}}=\mbox{Tr}_{A}\rho_{A\tilde{B}}$, that is $S(\rho_{\tilde{B}})=H_{\mbox{\tiny{bin}}}(\frac{1-F_-}{2})$.

\begin{figure}[hbtp]
\includegraphics[width=.45\textwidth]{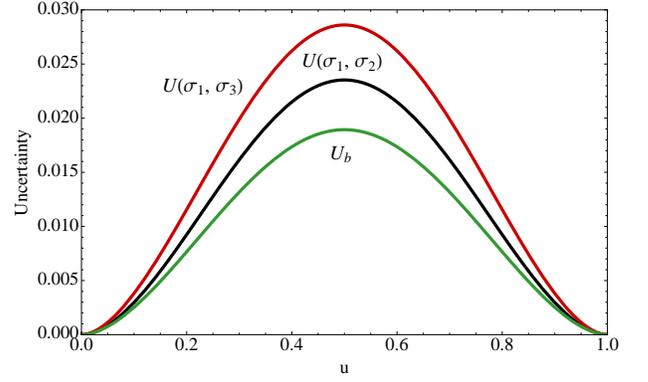}
\caption{The uncertainty of the measurement on observables $\sigma_1$ and $\sigma_3$ depends on the duration time of acceleration of Bob's cavity. The maximally entangled initial state with $c_1=-c_2=c_3=1$ and $k=1,\ s=0$ has been chosen, and $u=\eta_1/(2\ln(x_2/x_1))=h\tau_1/[4Lx_1\tanh{h/2}]$. To demonstrate the low acceleration approximation, the uncertainty is estimated under $h=0.1$.}
\label{eur-cavity}
\end{figure}

For a particular measurement of $\sigma_1$ and $\sigma_3$ by Alice, we can give the LHS of (\ref{eur2})
\bea
\tilde{U}(\sigma_1,\sigma_3)&=&H_{\mbox{\tiny{bin}}}(\frac{1-\tilde{\lambda}_1}{2})-2H_{\mbox{\tiny{bin}}}(\frac{1-F_-}{2})\no\\
&&+H'_{\mbox{\tiny{bin}}}(\tilde{\lambda}_3^+)+H'_{\mbox{\tiny{bin}}}(\tilde{\lambda}_3^-)+1
\label{cavityu}
\eea
which is depicted in Fig. \ref{eur-cavity} for a special choice of maximally entangled state.

It is interesting to note that the uncertainty is now periodic in time $\tau_1$, which measures the duration of the Bob's acceleration, with the period $T=4Lx_1\tanh(h/2)/h$. This is because the uncertainty depends on the entanglement generation between two cavity modes, which has been proved to be periodic \cite{BOX1,BOX2}. By properly choosing the parameters to ensure that $\tau_1=nT$ with $n\in \mathbb{N}$, the uncertainty bound is protected. On the other hand, for arbitrary duration, we demonstrate that the uncertainty change is very small, since the low acceleration approximation $|k|h\ll1$ has been imposed. Nevertheless, the range of our estimation is still under the ability of modern technology \cite{EUR3}, therefore could serve the future experiment test.

Finally, we calculate the RHS of the EUR (\ref{eur2}). The eigenvalues of (\ref{cavity1}) are
\bea
\tilde{\eta}_{1,\pm}&=&\frac{1}{4}\bigg[1+c_3(1-F_+)\pm\sqrt{F_-^2+(c_1-c_2)^2|G_k+\mathbb{A}^{(2)}_{kk}|^2}\bigg]\no\\
\tilde{\eta}_{2,\pm}&=&\frac{1}{4}\bigg[1-c_3(1-F_+)\pm\sqrt{F_-^2+(c_1+c_2)^2|G_k+\mathbb{A}^{(2)}_{kk}|^2}\bigg]\no
\eea
this give the RHS of (\ref{eur2}) which is
\be
\tilde{U}_b=-\sum_{i=1,2;
j=\pm}\tilde{\eta}_{i,j}\log_2\tilde{\eta}_{i,j}-H_{\mbox{\tiny bin}}(\frac{1-F_-}{2})+1
\label{cavityub}
\ee
also depicted in Fig. \ref{eur-cavity} for one of Bell states. Similar behavior as $\tilde{U}$ can be read. It is clear that the uncertainty in a game with two rigid cavities is always degraded. For any bipartite system without entanglement initially, e.g. $c_i=0$, we have $U_b=1$, recovering the RHS of (\ref{eur1}). We also conclude that, for integer periods $T$, $\tilde{U}=\tilde{U}_b$ is strictly satisfied in EUR (\ref{eur2}). Apparently, for the general state (\ref{bell}), this is extended to a SPMC condition $c_i=-c_jc_k\  (i\neq j\neq k)$ \cite{APP1}. However, for arbitrary duration $\tau_1$, this condition would be very complicated.

\section{Discussions}
\label{4}

In this Letter, we discuss the quantum-memory-assisted EUR in relativistic framework. We show how the Unruh effect for an accelerated observer increase the measurement uncertainty. In another uncertainty game, where both Alice and Bob localize in a rigid cavity respectively, therefore preventing the thermal noise from the acceleration horizon, we show that the uncertainty bound evolves periodically due to the cavity's motion can still affect the entanglement. By the properly chosen duration of the acceleration of cavity, the uncertainty bound can be protected. In both scenarios, we give the motion-dependent entropic uncertainty for the Bell-diagonal states, from which a SPMC condition can be derived. We utilize the low-acceleration approximation in cavity scenario, this assure that our results match the accuracy of current experimental technology.

We would like to mention some of the implications of our results. Firstly, it is well known that the uncertainty game with Unruh effect is equivalent to a similar process under Hawking radiation from a black hole, since the near-horizon geometry of Schwarzschild is described by same Rindler coordinates. Therefore our results could shed new light on the quantum physics near in a black hole background \cite{GUP}. Secondly, while the uncertainty in two-cavity scenario always increases, nevertheless, by a similar analysis, uncertainty for the game localized in a single cavity could be reduced, since the entanglement between different field modes is generated periodically \cite{BOX3}. For a bipartite system without entanglement initially, the EUR (\ref{eur1}) satisfied initially should be replaced by quantum-memory-assisted EUR (\ref{eur2}) in the final inertial segment III'. This scenario with single rigid cavity could be very instructive, according to equivalence principle, especially for the weakly gravitational background \cite{GRA1}. Since the entanglement between different spacetime positions can be extracted \cite{GRA2}, our analysis on the accelerated systems in Minkowski spacetime could be easily extended to curved background \cite{FUT1}, where the quantum-memory-assisted EUR is believed playing a fundamental role. It would be very interesting to investigate the implication of our results in quantum gravity theory \cite{FUT2}. Finally, we would like to comment on the nonlocality, along with uncertainty, characterizing the difference between classical and quantum physics. As the strength of quantum correlations might be radically altered in relativistic framework, it was shown that \cite{NLOC,BOX1} the nonlocality of system in terms of the violation of the CHSH inequality should also inherit the observer-dependence feature, while the rigidity of the Tsirelson bound is guaranteed by the so-called information causality \cite{ADD2,ADD3}. On the other hand, the nonlocality is inextricably and quantitatively linked to uncertainty \cite{ADD4} and can be determined uniquely by it. Consequently, the entropic uncertainty bound derived in (\ref{lhs1}) (\ref{cavityu}) could determine the nonlocality of the system \cite{ADD5}. Generalized to a dynamic spacetime (e.g. Robertson-Walker space), this link between nonlocality and entropic uncertainty should be particularly powerful in understanding the emergence of classical behavior in quantum system.

\section*{Acknowledgement}
This work is supported by NSFC, 973 program through 2010CB922904. J. F. thanks Yu-Ran Zhang and Kai Zhang for stimulating discussions. Y. Z. Z. and M. D. G. acknowledge the support of the Australian Research Council through DP110103434.

\appendix

\section{}
\label{5}
Here we summarize the calculation of the coefficients in (\ref{coef}). Introducing the abbreviations
\[
F_\pm=f^+_k\pm f^-_k=\sum_{p\geqslant0}|\mathbb{A}^{(1)}_{pk}|^2\pm\sum_{p<0}|\mathbb{A}^{(1)}_{pk}|^2\no
\]
$F_+$ is given in \cite{BOX1}
\bea
F_+&=&\sum_{p=-\infty}^{\infty}|E_1^{k-p}-1|^2|A^{(1)}_{kp}|^2\no\\
&=&\frac{4h^2}{\pi^4}[4(k+s)^2(Q_6(1)-Q_6(E_1))+Q_4(1)-Q_4(E_1)]\no
\eea
where we use the notation $Q_\alpha(\beta)\equiv\mbox{Re}\big[\mbox{Li}_\alpha(\beta)-\frac{1}{2^\alpha}\mbox{Li}_\alpha(\beta^2)\big]$, Li is the polylogarithm and $E_1\equiv\exp(\frac{i\pi\eta_1}{\ln(x_2/x_1)})=\exp(\frac{i\pi h\tau_1}{2Lx_1\tanh(h/2)})$.

$F_-$ can be calculated similarly as 
\bea
F_-&=&\bigg(\sum_{p\geqslant0}-\sum_{p<0}\bigg)|E_1^{k-p}-1|^2|A^{(1)}_{kp}|^2\no\\
&=&\frac{16h^2}{\pi^4}2(k+s)[Q_5(1)-Q_5(E_1)]+P(k,s,E_1)\no
\eea
where $P$ is a polynomial summing for all terms with odd number
\[
\sum_{m=1}^{k}\frac{4h^2}{\pi^4}\big(1-\mbox{Re}(E_1^{m})\big)\bigg[4(k+s)\bigg(\frac{k+s}{m}-1\bigg)+\frac{1}{m^4}\bigg]
\]
For instance, $P(0,s,E_1)=0$. For $k=1$, it gives $P=\frac{4h^2}{\pi^4}\big(1-\mbox{Re}(E_1)\big)\big[4s(1+s)+1\big]$.

From the relation between $A$ and $\mathbb{A}$ given in \cite{BOX1}, we have 
\bea
\mathbb{A}^{(2)}_{kk}&=&-E_1^{k+s}h^2\bigg\{\bigg(\frac{1}{48}+\frac{\pi^2(k+s)^2}{120}\bigg)\no\\
&&-\frac{2}{\pi^4}\big[4(k+s)^2Q_6(E_1)+Q_4(E_1)\big]\bigg\}\no
\eea
which gives
\bea
|G_k+\mathbb{A}^{(2)}_{kk}|^2&=&\bigg(1-h^2\bigg\{\bigg(\frac{1}{48}+\frac{\pi^2(k+s)^2}{120}\bigg)\no\\
&&-\frac{2}{\pi^4}\big[4(k+s)^2Q_6(E_1)+Q_4(E_1)\big]\bigg\}\bigg)^2\no
\eea

\end{document}